\documentclass[12pt,a4paper]{article}
\usepackage[dvipdfmx]{color}
\usepackage[dvipdfmx]{graphicx}

\usepackage{float}
\usepackage{amsfonts}
\usepackage{amssymb,amsmath}
\usepackage{latexsym}
\input{colordvi.tex}

\usepackage{cite}
\usepackage[normalem]{ulem}

\renewcommand{\thefootnote}{\#\arabic{footnote}}

\begin{document}

\newcommand{\TT}[1]{\Red{{\bf #1}}}
\newcommand{\TS}[1]{\Green{{\bf [#1]}}}

\renewcommand{\thepage}{\arabic{page}}
\setcounter{page}{1}
\renewcommand{\thefootnote}{\#\arabic{footnote}}

\begin{titlepage}

\begin{center}

\vskip .5in

{\Large \bf Mixed Inflaton and Spectator Field Models: \\
CMB constraints and $\mu$ distortion 
}

\vskip .45in

{\large
Kari~Enqvist$\,^{1}$, Toyokazu~Sekiguchi$\,^{1,2}$ and Tomo~Takahashi$\,^3$
}

\vskip .45in

{\em
$^1$
Department of Physics and
Helsinki Institute of Physics, FIN-00014, \\
 University of Helsinki, Finland
\\
$^2$
Institute for Basic Science, Center for Theoretical Physics of the Universe, \\
Daejeon 34051, South Korea
\\
$^3$
Department of Physics, Saga University, Saga 840-8502, Japan
}

\end{center}

\vskip .4in

\begin{abstract}

We discuss mixed inflaton and spectator field models where both the fields 
are responsible for the observed density fluctuations.  We use the current CMB data to constrain both the general mixed model as well as some specific representative scenarios, and collate the results with the model predictions for the CMB spectral $\mu$ distortion.
We find the posterior distribution of $\mu$ using MCMC chains and demonstrate that the standard single-field inflaton model typically predicts $\mu \sim 10^{-8}$ with a relatively narrow distribution, whereas for the mixed models, the distribution turns out to be much broader, and $\mu$ could be larger by almost an order of magnitude. 
Hence future experiments of $\mu$ distortion could provide a  tool for the critical testing of the mixed source models of the primordial perturbation.

\end{abstract}
\end{titlepage}

\setcounter{footnote}{0}

%%%%%%%%%%%%%%%%%%%%%%%
\section{Introduction}
%%%%%%%%%%%%%%%%%%%%%%%

Quantum fluctuations of the inflaton are usually considered as the main contender for explaining the origin of the primordial curvature perturbation, with a host of 
inflaton models in existence \cite{Martin:2013tda}. As is well known, the observed density perturbations of the cosmic microwave background (CMB)  could also be generated by an effectively massless  scalar  field 
other than the inflaton such as the curvaton \cite{Enqvist:2001zp,Lyth:2001nq,Moroi:2001ct}, or by scalars in the
modulated reheating scenarios  \cite{Dvali:2003em,Kofman:2003nx}. 
In those models, the inflationary expansion is driven by the inflaton field while the energy density of the other scalar(s) is not dynamically important during inflation. Such a field is therefore called a ``spectator field."

However, the effectively massless spectators are always subject to quantum fluctuations in the inflationary Universe.
At a later time these may be imprinted on the metric as a curvature perturbation. Since the inflaton also acquires perturbations, in general the primordial fluctuations could be sourced both by the inflaton and some spectator fields.
Such a possibility is sometimes referred as the mixed inflaton and spectator field model\footnote{
Such mixed models have been 
investigated in the curvaton case in \cite{Langlois:2004nn,Lazarides:2004we,Moroi:2005kz,Moroi:2005np,Ichikawa:2008iq,Fonseca:2012cj,Enqvist:2013paa,Vennin:2015vfa}, in modulated reheating \cite{Ichikawa:2008ne}  as well as in the case of a general spectator field \cite{Suyama:2010uj,Enqvist:2013paa}. 
}. The issue at hand is then: how to test these different possibilities.

At the moment the most powerful tool for probing the primordial perturbation is provided by the precision observations of the CMB anisotropies by the  Planck satellite, which constrain possible models of inflation and/or inflaton potentials  \cite{Ade:2015lrj}. 
However, in terms of inflationary efolds $N$, the information thus obtained pertains a rather narrow range $\Delta N\sim {\cal O}(5)$ or so. This makes it difficult to determine spectral properties of the primordial perturbation on broad-scales, e.g., 
the running of the spectral index, and to differentiate theoretical models.

In the near future there may however open up a new observational window on the primordial perturbation that could help to test models for the origin of the curvature perturbation over a much wider range of efolds. The basic idea is that the original CMB spectrum will generally obtain a very small deviation from the perfect black body by virtue of imperfect thermalization of the energy dissipated from density fluctuations. At early times the energy released into photons via Silk damping is thermalized rapidly by the single and double Compton scattering. However, at late times the double Compton scattering becomes less effective. As a result, released energy is imprinted on the CMB spectrum as distortions, creating effectively a chemical potential when the kinetic equilibrium is maintained by the single Compton scattering. This is the so-called $\mu$ distortion of the CMB, which can be detected in the future  \cite{Kogut:2011xw,Andre:2013nfa} with a sensitivity by several orders of magnitude better than the current ones (i.e. COBE FIRAS \cite{Mather:1993ij,Fixsen:1996nj}) and some studies have been performed focusing on the probing primordial fluctuations using the $\mu$ distortion \cite{Chluba:2012we,Dent:2012ne}.

In general, the power spectra of fluctuations from different scalar fields exhibit different scale dependences.
Thus it is possible that, for example,  fluctuations sourced by one field dominate on large scales, while 
those from another scalar field can give a significant contribution on smaller scales. 
If this were the case, the prediction of a model with only one scalar field could 
become inconsistent when the large and small scale observations are compared. Such considerations could thus provide the means to differentiate between the single field and mixed models\footnote{
When there is some modulation in the potential of the inflaton, or some other field responsible for the density fluctuations, 
a deviation from a smooth potential could also be a possibility \cite{Enqvist:2014nsa}.
}. This is where the determination of the $\mu$ distortion of the CMB can play a decisive role.

In the present paper we investigate models where the curvature perturbation has two sources.  Although our treatment is generic, we have mostly in mind mixed inflaton and spectator field models. We find constraints on such mixed source models from the current CMB observations and study the prospects for testing and constraining mixed models by the $\mu$ distortion of CMB .

The paper is organized as follows. 
In the next section, we introduce the mixed source model to be investigated
and summarize the formalism needed for the studies in the subsequent sections.
We also briefly summarize the procedure to calculate the $\mu$
distortion. Then in Section~\ref{sec:const}, we present our results on the constraints on  mixed source models 
implied by current CMB data and compute the predictions for the $\mu$ distortion in terms of the model parameters.  
The final section is devoted to the summary and conclusions.

%%%%%%%%%%%%%%%%%%%%%%%
\section{Formalism}
\label{sec:model}
%%%%%%%%%%%%%%%%%%%%%%%

%%%%%%%%%%%%%%%%%%%%%%%
\subsection{Mixed model}
%%%%%%%%%%%%%%%%%%%%%%%

Let us consider a model with two sources of density fluctuations with 
different scale dependences. In general, when there are two sources of density fluctuations, 
the primordial power spectrum  can be given as their sum:
\begin{equation}
\label{eq:power_12}
P_\zeta(k) 
= P_\zeta^{(1)} + P_\zeta^{(2)}
= A_{s1} (k_{\rm ref}) \left( \frac{k}{k_{\rm ref}} \right)^{n_{s1} -1}
+ A_{s2} (k_{\rm ref}) \left( \frac{k}{k_{\rm ref}} \right)^{n_{s2} -1},
\end{equation}
where $A_{s1}$ and $A_{s2}$ are the amplitude of fluctuations for the source 1 and 2, respectively. 
Here $n_{s1}$ and $n_{s2}$ are the spectral indices for each source, while 
$k_{\rm ref}$ is the reference wave number at which the amplitude and the spectral index are 
measured. Unless otherwise stated,  in this paper we take $k_{\rm ref} =0.05~{\rm Mpc}^{-1}$.
To represent the relative amplitude between these two power spectra, we define 
\begin{equation}
\label{eq:def_R}
R \equiv  \frac{ P_\zeta^{(2)} (k_{\rm ref}) }{P_\zeta^{(1)} (k_{\rm ref})},
\end{equation}
which is evaluated at the reference scale.

The specific model we have in mind is the mixed inflaton and spectator model. 
There  the power spectrum is sourced by both the inflaton $\phi$ and the spectator field $\sigma$. 
The primordial power spectrum due to the inflaton is given by
\begin{equation}
\label{eq:power_inf}
P_\zeta^{(\phi)}  = \frac{1}{2 \epsilon M_{\rm pl}^2} \left( \frac{H_{\rm inf}}{2\pi} \right)^2,
\end{equation}
where $\epsilon$ is a slow-roll parameter defined below (see Eq.~\eqref{eq:slow_roll}), 
$H_{\rm inf}$ is the Hubble parameter during inflation 
and $M_{\rm pl}$ is the reduced Planck energy scale.

If we assume that the spectator field $\sigma$ is the curvaton, the power spectrum can be written as
\begin{equation}
\label{eq:power_cur}
P_\zeta^{(\sigma)}  = \left( \frac{2 r_{\rm dec}}{3\sigma_\ast} \right)^2  \left( \frac{H_{\rm inf}}{2\pi} \right)^2,
\end{equation}
where $r_{\rm dec}$ roughly denotes the energy fraction of the curvaton to the total one at its decay, 
 which is defined by 
\begin{equation}
r_{\rm dec} = \left. \frac{3\rho_\sigma}{3 \rho_\sigma + 4 \rho_r } \right|_{\rm dec},
\end{equation}
while $\sigma_*$ denotes the value of $\sigma$ at the time of horizon exit.

The spectral indices of the power spectra for the inflaton and the curvaton can respectively be written as \cite{Byrnes:2006fr,Kobayashi:2013bna}\footnote{
Although we give the expression of the power spectrum for the curvaton case in Eq.~\eqref{eq:power_cur},  
the following expression for $n_s^{(\sigma)}$ is valid for any isocurvature field during inflation.
}
\begin{equation}
\label{eq:ns_phi_sig}
n_s^{(\phi)} -1 = -6 \epsilon + 2\eta_\phi,
\qquad
n_s^{(\sigma)} -1 = -2  \epsilon + 2\eta_\sigma.
\end{equation}
In the above formulas, the slow-roll parameters are defined by 
\begin{equation}
\label{eq:slow_roll}
\epsilon = -\frac{\dot{H}}{H^2}, 
\qquad 
\eta_\phi = \frac{V_{\phi\phi}}{3 H^2},
\qquad
\eta_\sigma = \frac{V_{\sigma\sigma}}{3 H^2},
\end{equation}
with a dot representing the time derivative,  $V_{\phi\phi}$ and $V_{\sigma\sigma}$ 
being $V_{\phi\phi} = d^2 V / d\phi^2$, $V_{\sigma\sigma} = d^2 V / d\sigma^2$, respectively.

The tensor power spectrum is given as in the standard case by
\begin{equation}
%\label{ }
P_T = \frac{8}{M_{\rm pl}^2} \left( \frac{H_{\rm inf}}{2\pi} \right)^2.
\end{equation}
The tensor spectral index can be  written as $n_T = -2 \epsilon$. 

To discuss the size of the tensor mode, one usually uses the tensor-to-scalar ratio which 
 for the mixed model is given as
\begin{equation}
\label{eq:mixed_r}
r \equiv  \frac{P_T(k_{\rm ref})}{P_\zeta (k_{\rm ref})} \equiv \frac{16 \epsilon}{1+R}.
\end{equation}
Here $R$ is the measure of the contribution of the spectator, which has been defined in Eq.~\eqref{eq:def_R}.

When one compares the mixed model with cosmological observations such as CMB, 
one needs to calculate CMB angular power spectra for each primordial power spectrum separately, and
then add those contributions. However, when the scale dependences of the two power spectra are not that much different, 
which is usually the case, we can define an ``effective" spectral index for the summed power spectrum as
\begin{eqnarray}
\label{eq:ns_mix}
n_s^{\rm (eff)} - 1 &\equiv&  \frac{d }{d \ln k} \left\{ \ln \left( P_\zeta^{(\phi)} + P_\zeta^{(\sigma)} \right) \right\} \notag \\
&=&\frac1{1+R} \left( n_{s1} - 1 \right) + \frac R{1+R} \left( n_{s2} -1 \right) \label{eq:neff}\\
&=&- 2\epsilon + 2 \eta_\sigma + \frac{-4 \epsilon + 2 \eta_\phi - 2 \eta_\sigma}{1+R}.\notag
\end{eqnarray}
Then we can regard the sum of the power spectra as if it were a single spectrum with the above effective spectral index, and 
compare directly with observations. 
However, it should be noted that when the scale dependences of the two spectra are very different, 
this description is not valid. One then has to calculate the CMB power spectra separately and add them together, whence
 the constraints on the mixed models turn out to be quite non-trivial. We will investigate this case in Section \ref{sec:const}.

In closing this section, let us comment on the running of the spectral indices.
In general, the running is non-zero and such running may affect the power spectra considerably,  in particular  on small scales. However, since we focus on mixed source models of primordial fluctuations, 
the effect of the running would be subdominant as compared to that of the mixed-source nature. This should remain 
a good approximation especially for the setups  to be discussed in the following sections.

%%%%%%%%%%%
\subsection{CMB $\mu$ distortion}
%%%%%%%%%%%

In this subsection,  let us summarize the formalism to calculate the CMB $\mu$ distortion due to the energy dissipation 
generated by the acoustic waves, particularly from the viewpoint of constraining the primordial power spectrum.
We just give here  some key formulas.
For details, we refer the readers to \cite{Hu:1994bz,Chluba:2011hw,Chluba:2012we,Dent:2012ne,Chluba:2012gq}. We basically follow the argument of \cite{Dent:2012ne}.

When there is an energy injection $Q$, 
the distribution function of photon can deviate  from the black body form. Although such a deviation can be
characterized by several types of distortions such as $y$- and $i$-types \cite{Zeldovich:1969ff,Khatri:2012tw}, here we focus on the $\mu$-type distortion. 

For any given model, the spectral distortion can in principle be obtained numerically as a function of the model parameters.
Given the energy injection $Q$, $\mu$ can be calculated by \cite{Hu:1994bz} 
\begin{equation}
\label{eq:mu_evolv}
\frac{d\mu}{dt}  \simeq  - \frac{\mu}{t_{\rm DC} (z)} + \frac{1}{\rho_\gamma} \frac{dQ}{dt},
\end{equation}
where $\rho_\gamma$ is the energy density of photon and $t_{\rm DC} (z)$ is the double Compton time scale:
$
t_{\rm DC} (z) = 2.06 \times 10^{33}  \left( 1 - Y_P/2 \right)^{-1} (\Omega_b h^2)^{-1} z^{-9/2}~{\rm sec}.
$
$Y_p$ represents the Helium abundance, $\Omega_b$ is baryon density parameter and $z$ is the redshift.
Since here we consider the energy injection due to the dissipation of the acoustic waves, $Q$ is given by
$Q \simeq (\rho_\gamma /2) \left(  c_s^2 \left\langle  \delta_\gamma ^2 \right\rangle + \left\langle v_\gamma^2 \right\rangle \right)$
with $\delta_\gamma$ and $v_\gamma$ being density fluctuations and velocity of photons, and $c_s$ being the sound speed of photon fluid.
By using the transfer functions for $\delta_\gamma$ and $v_\gamma$, one can relate the energy injection $Q$ with the primordial power spectrum of photon $P_\gamma (k)$ 
as
$Q = \Delta_Q (k)^2 P_\gamma (k)$,  where $\Delta_Q = (3c_s)/\sqrt{2} e^{-k^2 /k_D^2}$.
Here $k_D$ is the diffusion length, which is given by
\begin{equation}
\frac{1}{k_D^2} = \int_z^{\infty} d\tilde{z} \frac{1+\tilde{z}}{6 H(\tilde{z}) (1+R_{b\gamma}) n_e \sigma_T} 
\left( \frac{R_{b\gamma}^2}{1+R_{b\gamma}} + \frac{16}{15} \right),
\end{equation}
where $H(z)$ is the Hubble parameter, $n_e$ is the number density of free electron, 
$\sigma_T$ is the Thomson scattering cross section and $R_{b\gamma} = 3\rho_b / (4\rho_\gamma )$.

Since the primordial (curvature) power spectrum $P_\zeta (k)$ is related to the photon power spectrum $P_\gamma (k) $ 
as $P_\gamma (k) = A P_\zeta (k)$ with $A = 4 /(2R_\nu /5 + 3/2)^2$ and $R_\nu = \rho_\nu /(\rho_\gamma + \rho_\nu)$ being the fraction of the neutrino energy 
density, the dissipation energy can be given once 
we specify the primordial power spectrum. (We also need to specify some cosmological parameters such as the baryon density, the Helium abundance and so on;
however the dependence of $\mu$ on those parameters is very weak.)
  By inserting all this information into Eq.~\eqref{eq:mu_evolv} and integrating, 
we obtain
\begin{equation}
\mu = 1.4 \int_{z1}^{z2} dz e^{-z^2 /z_{\rm DC}^2} \left( - \int \frac{dk^3}{(2\pi)^3} A P_\zeta (k) \frac{\Delta_Q^2}{dz} \right),
\end{equation}
where $z_{\rm DC} = 4.1 \times 10^5 (1 - Y_p /2)^{-2/5} (\Omega_bh^2)^{-2/5}$ 
represents the redshift of the time scale for the double Compton scattering.

The generation of the $\mu$ distortion is effective during 
when $z_2  <  z  < z_1$ with $z_1= 2 \times 10^6$ and $z_2 = 5 \times 10^4$, which set the upper and lower bounds of the integral.
By using the power spectrum given in Eq.~\eqref{eq:power_12}, we can calculate the $\mu$ distortion for any parameter choice of the mixed inflaton and 
spectator field model. Note that the $\mu$ distortion does not depend on any additional parameters (except  weak dependences on some parameters mentioned above) 
but is determined completely by the primordial spectra.

Current bound on the $\mu$ distortion  is $ \mu < 9\times 10^{-5} ~({\rm 95 \%~ C.L.})$, which 
has been obtained from COBE FIRAS \cite{Mather:1993ij,Fixsen:1996nj}\footnote{
Another bound has been obtained from ARCADE, which gives $ \mu < 6\times 10^{-4} ~({\rm 95 \%~ C.L.})$ \cite{ARCADE}.
}.
However, in the future, the sensitivity to $\mu$ will be much improved.
Future observations such as PIXIE and PRISM are expected to probe the distortion at the level of 
$\mu \sim  10^{-8}$ and $10^{-9}$, respectively  \cite{Kogut:2011xw,Andre:2013nfa}.

%%%%%%%%%%%%%%%%%%%%%%%
\section{Current CMB constraints and the $\mu$ distortion}
\label{sec:const}
%%%%%%%%%%%%%%%%%%%%%%%

Let us now discuss the current CMB constraints on a mixed model combined with
the predictions for the $\mu$ distortion. 
To obtain observational constraints from CMB, we adopt the data from Planck \cite{Ade:2013kta}\footnote{
Here we use the data from Planck 2013. Although the data from Planck 2015 is now available, 
the constraints should turn out to be almost the same so that the results presented in this paper should remain unchanged.
}, BKP (joint analysis of BICEP2/Keck and Planck) \cite{Ade:2015tva} and  CMB high $\ell$ data from 
Atacama Cosmology Telescope (ACT) and  South Pole Telescope (SPT)  \cite{Dunkley:2013vu,Keisler:2011aw,Reichardt:2011yv}.
We perform  Markov Chain Monte Carlo (MCMC) analysis with a modified version of  {\tt cosmomc} code \cite{Lewis:2002ah}.

In the following, we focus on constraints on the parameters describing primordial density fluctuations. 
From the point  of view of the analysis, the simplest way to describe the primordial power spectra for the mixed models would be to  
use the amplitudes of the two power spectra $A_{s1}$ and $ A_{s2}$ 
(or $A_{s1}$ and the ratio $R=A_{s2}/A_{s1}$) 
and their spectral indices $n_{s1}$ and $n_{s2}$ at 
the given reference scale. 
On the other hand, from a theoretical view point, 
it may be useful to see the constraints on the slow-roll parameters ($\epsilon, \eta_1, \eta_2$). Thus, in presenting our results, 
we show the constraints for both of these parametrizations, which we describe in detail below.
We note that the tensor-to-scalar ratio $r$ is also varied unless otherwise stated. 
In addition to the parameters characterizing the primordial power spectra, 
we also vary  the standard cosmological parameters:
baryon energy density $\omega_b = \Omega_bh^2$, dark matter energy density $\omega_c = \Omega_ch^2$, 
the acoustic peak scale $\theta_s$ and reionization optical depth $\tau$.

Our results are presented below. Since we have performed analyses with several setups, in which 
the treatments of the parameters differ, we display our results for each case separately.

\bigskip
\bigskip
\noindent 
{\bf (i) Standard case (pure inflaton case)}

\begin{figure} %[htbp]
  \begin{center}
    \resizebox{160mm}{!}{
%    \hspace{-20mm}
    \includegraphics{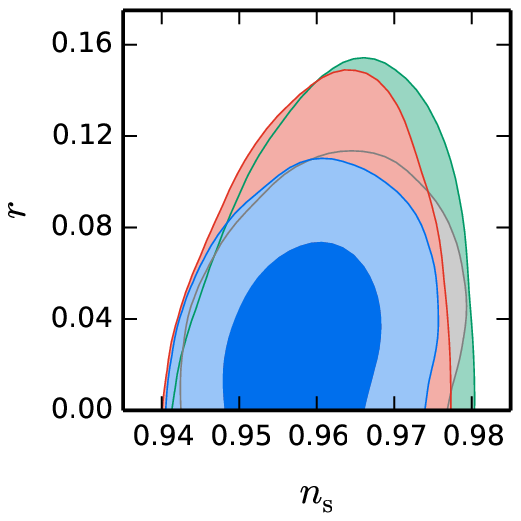}\hspace{3mm}
    \includegraphics{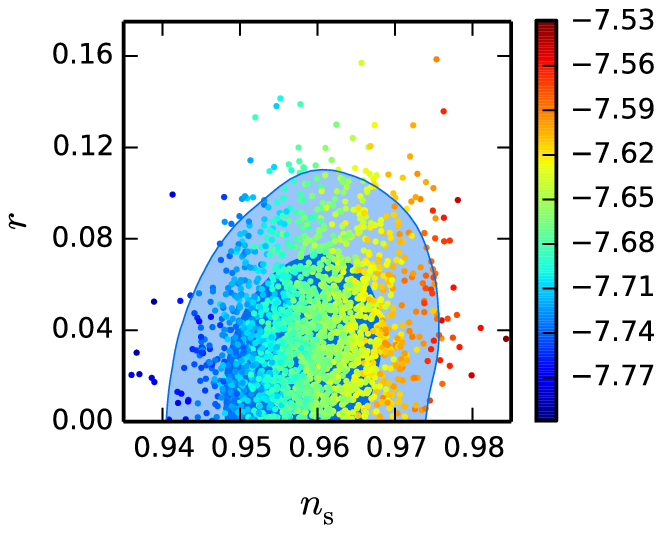}
    }
  \caption{The standard inflation case (no spectator field):
  [Left] Constraints on the $n_s$--$r$ plane from Planck alone (green), Planck+high $\ell$ data (gray), 
  Planck+BKP (red) and  Planck+BKP+high $\ell$ data (blue).  1$\sigma$ and 2$\sigma$ constraints are shown with dark and light colors, respectively.
  [Right] Scatter plot of $\mu$ whose values are indicated in the color legend. For reference, the constraint from Planck+BKP+high $\ell$ data (blue region) is also shown.}
     \label{fig:ns_r_inf}
  \end{center}
  \end{figure}

\bigskip
First we show the results for the standard case where primordial fluctuations are generated solely from 
the inflaton, i.e. $R=0$.
 Although the $\mu$ distortion for the standard case has already been investigated in detail, including the running of the spectral index, 
in Refs.~\cite{Chluba:2012we,Chluba:2012we,Dent:2012ne},  we show  here the results for the benefit of comparison with the mixed model to be discussed subsequently.
In Fig.~\ref{fig:ns_r_inf}, we display the constraint from current CMB data and a scatter plot of the predicted value for $\mu$ 
in the $n_s$--$r$ plane on the left and right panels, respectively. 
CMB constraints are shown from the analysis employing Planck alone (green region), Planck+high $\ell$ data (gray region), 
Planck+BKP (red region) and Planck+BKP+high $\ell$ data (blue region). 1$\sigma$ and 2$\sigma$ allowed regions are 
given with dark and light colors.

Although we show a scatter plot of $\mu$ in the $n_s$--$r$ plane, to calculate the value of $\mu$, 
we also need to fix the amplitude $A_s$. However, the amplitude is well determined by CMB observations, and
thus it can be regarded as almost fixed. Once the amplitude is fixed, the value of $\mu$ can be evaluated just by giving the spectral index $n_s$\footnote{
In fact, the tensor mode can also generate $\mu$ distortion \cite{Ota:2014hha,Chluba:2014qia}. However, the distortion can be comparable to 
the one generated by the scalar mode only when the tensor spectral index is blue-tilted  \cite{Ota:2014hha,Chluba:2014qia}. Therefore here we neglect such a contribution and hence
the tensor-to-scalar ratio is here irrelevant for calculating the value of $\mu$.
}.   Thus the region with larger $\mu$ corresponds to the one with larger $n_s$, as seen from the right panel of Fig.~\ref{fig:ns_r_inf}.
In the standard inflation case, given the current CMB constraints, $\mu$ typically takes a value $\mu \sim 2 \times 10^{-8}$, 
which may be detectable in the future observations such as PIXIE and PRISM.

\bigskip
\bigskip
\noindent 
{\bf (ii) General mixed model  }

\bigskip
In this case, we  vary the parameters $A_{s1}, R, n_{s1}$ and $n_{s2}$ freely, 
which corresponds to the general two-source case. In general, the parameter $R$ which is related to the scalar-to-tensor ratio can take any value.
However, $R$ and the spectral indices $n_{s1}$ and $n_{s2}$ are highly degenerate, which worsens the convergence of the MCMC chain.
Thus to avoid such a degeneracy,
 in our analysis we assume a finite prior range for $R$ as $10^{-2} \le R \le 10^2$. The effect of this prior on the results are discussed below.

\begin{figure} %[htbp]
  \begin{center}
    \resizebox{160mm}{!}{
    \includegraphics{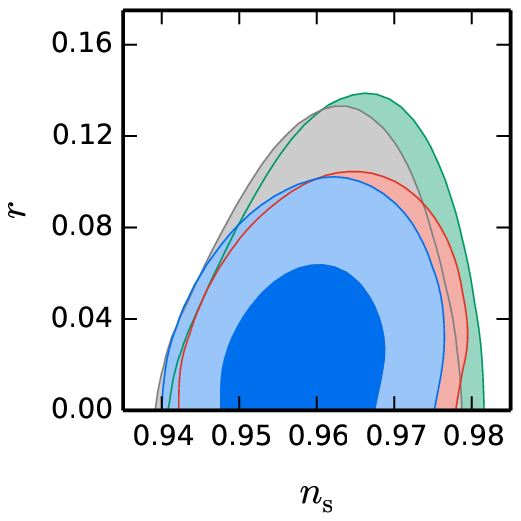}\hspace{3mm}
    \includegraphics{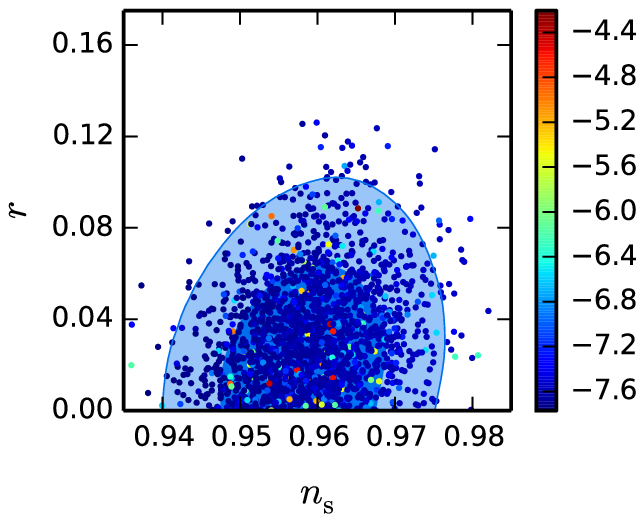}
    }
\caption{The same as Fig.~\ref{fig:ns_r_inf}, but for the general mixed source case. Notice that  
  the spectral index $n_s$ here is the ``effective" spectral index  defined in Eq.~\eqref{eq:neff}.}
     \label{fig:ns_r_mixed}
  \end{center}
  \end{figure}

In Fig.~\ref{fig:ns_r_mixed}, we show respectively the CMB constraint and a scatter plot of $\mu$ for the case with a general mixed model on the left and right panels in 
the $n_s^{\rm (eff)}$--$r$ plane, where $n_s^{\rm (eff)}$ is the effective spectral index defined in Eq.~\eqref{eq:neff}.
It should be noted that we introduced $n^{\rm (eff)}$  just for illustrative purposes
and the total power spectrum cannot be described by a simple power-law form.
As one can see from the right panel,  in this kind of a mixed source model , the value of $\mu$ can be as large as $10^{-5}$ (or larger)
even inside the allowed region for certain parameter values. Notice that, although $n_s^{\rm (eff)}$ should be around  0.96, the spectral indices $n_{s1}$ and $n_{s2}$ can be respectively
very blue-tilted when $R \gg 1$ and $R \ll 1$.
 
\begin{figure} %[htbp]
  \begin{center}
    \resizebox{160mm}{!}{
    \includegraphics{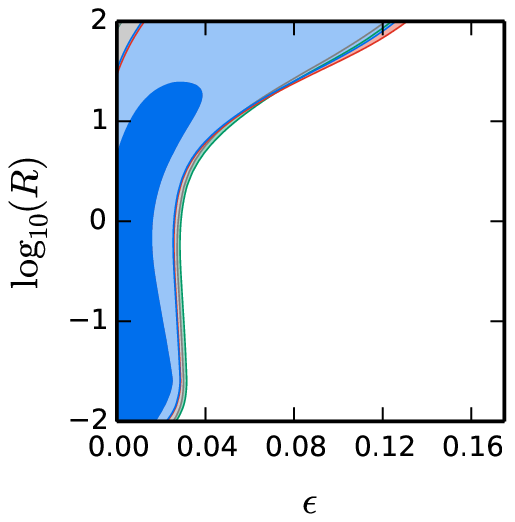}\hspace{3mm}
    \includegraphics{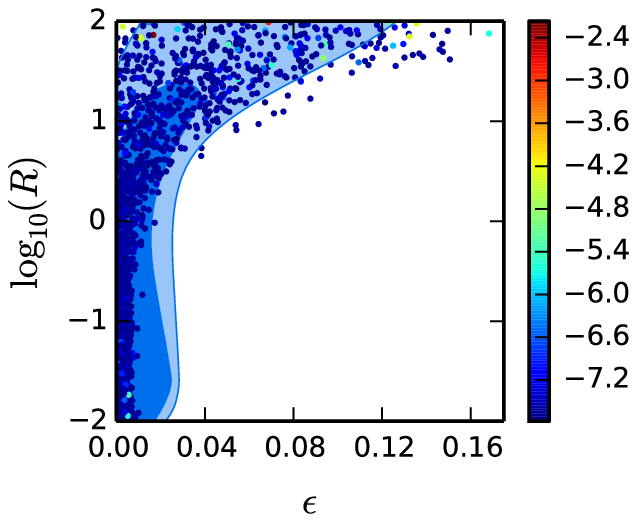}
    }
  \caption{CMB constraints and a scatter plot of $\mu$ on the $\epsilon$--$R$ plane for the general mixed case.
  The constraints are represented with the same color regions with those in Fig.~\ref{fig:ns_r_mixed}.}
     \label{fig:eps_R_mixed}
  \end{center}
  \end{figure}

To show that $R$ is scarcely constrained by the current CMB observations, we depict the CMB constraint on 
the $\epsilon$--$R$ plane in the left panel of Fig.~\ref{fig:eps_R_mixed}. A scatter plot for the value of $\mu$ is also 
shown in the right panel. Notice that the tensor-to-scalar ratio for the general mixed model is given by Eq.~\eqref{eq:mixed_r}.
Since current observations just give an upper bound on $r$,  we cannot obtain any bound on $R$. 
Regardless of the value of $\epsilon$,
a very large value of $R$ can be consistent with the current observational bounds on $r$ since 
the tensor-to-scalar ration is suppressed in this case.
This also explains the fact that a broader range of $\epsilon$ is allowed in the large $R$ region. 
Small values of $R$ are also acceptable since such cases simply reduce to the standard inflaton case.

\begin{figure} %[htbp]
  \begin{center}
    \resizebox{160mm}{!}{
    \includegraphics{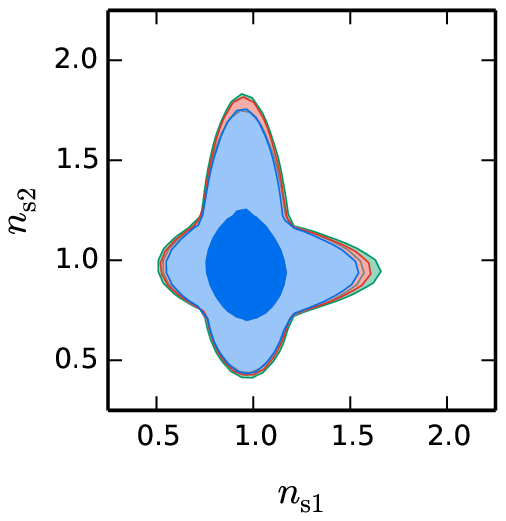}
    \includegraphics{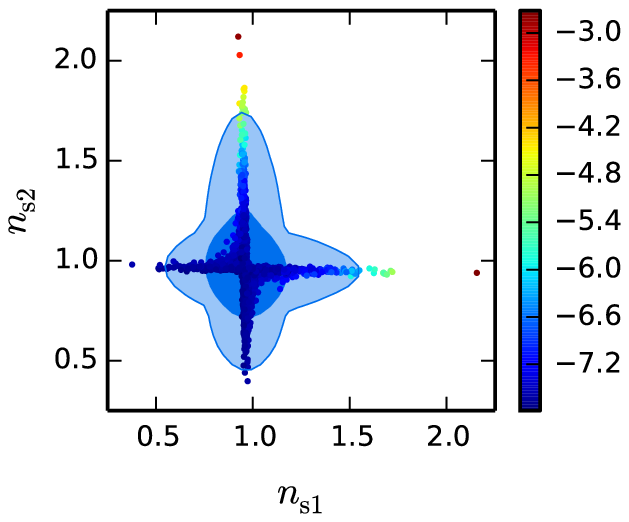}
    }
  \caption{CMB constraints and a scatter plot of $\mu$ on the $n_{s1}$--$n_{s2}$ plane for the general mixed case. 
  Note that the allowed region in the $n_{s1}$--$n_{s2}$ plane depends on the prior on $R$. See the text for more explanation.}
     \label{fig:ns1_ns2_mixed}
  \end{center}
  \end{figure}

To see which values of $n_{s1}$ and $n_{s2}$ would correlate with  large values of $\mu$, we also show our result in the $n_{s1}$--$n_{s2}$ plane as well in 
Fig.~\ref{fig:ns1_ns2_mixed}. We should note here that, although there seem to appear upper bounds on $n_{s1}$ and $n_{s2}$, 
they depend on the prior for $R$.  As explained above, due to computational reasons,
we adopted a prior for $R$ as $10^{-2} \le R \le 10^2$. However, if we instead were to assume a broader range, for example, 
$10^{-3} \le R \le 10^3$,  the allowed region for the spectral indices would be broadened.  One can see that 
the points with large values of $\mu$ lie in the region where either spectral index is very blue-tilted.

\bigskip
\bigskip
\noindent
{\bf (iii) Mixed models with $\eta_2=0$}

\begin{figure} %[htbp]
  \begin{center}
    \resizebox{160mm}{!}{
    \includegraphics{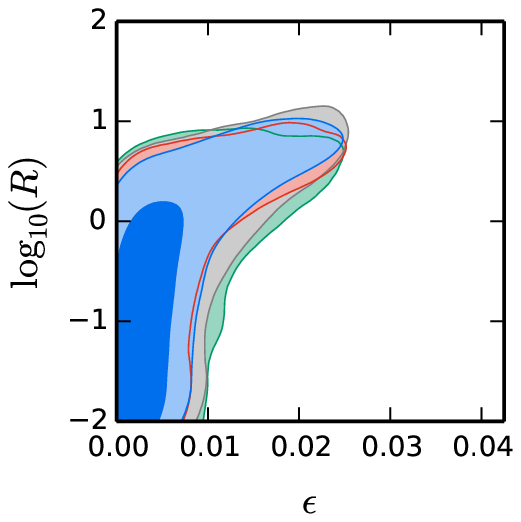}\hspace{3mm}
    \includegraphics{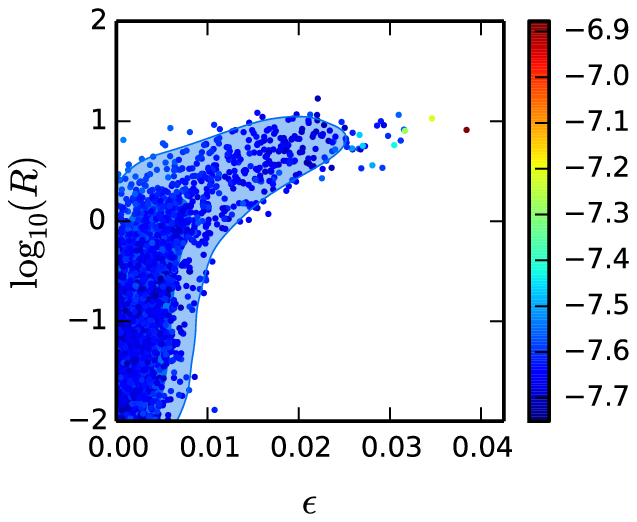}
    }
  \caption{The same as Fig.~\ref{fig:eps_R_mixed}, but for the case with $\eta_2=0$. Some caution is necessary when interpreting the 
  constraint from CMB in this case. See the text for the detail.}
     \label{fig:eps_R_eta0}
  \end{center}
  \end{figure}

\bigskip
In many cases, the mass of the spectator field is assumed to be very light as compared 
to the inflationary Hubble scale, whence $\eta_2$ should be very small. 
Therefore, here we consider such a situation by fixing $\eta_2=0$. 
In Fig.~\ref{fig:eps_R_eta0}, we show the CMB constraints and the value of $\mu$ in the same manner of
Fig.~\ref{fig:eps_R_mixed}.

Although it seems that there is an upper bound on $R$ in this case, some caution is necessary when interpreting this 
result. Since $\eta_2 =0$ is fixed, the spectral index for the spectator $n_{s2}$ is given by $n_{s2} -1 = - 2 \epsilon$. 
When $R$ is large, the power spectrum is dominated by the contribution from $P_\zeta^{(2)}$, and
$n_{s2}$ should be close to $0.96$ (roughly the mean value from Planck result \cite{Ade:2015lrj}).
This fixes the value of $\epsilon$ as $\epsilon \simeq 0.02$.
When $\epsilon$ is fixed, $r$ and $R$ are related by Eq.~\eqref{eq:mixed_r} as $16\epsilon = r (1+R)$,
which indicates that $r$ should be very small when $R$ is large.  In particular, when $R \sim10^2$, which is the upper edge of the prior on $R$, 
the tensor-to-scalar ratio should be small as $r \sim 10^{-3}$. The region with such a small value of $r$ tends to be difficult to be sampled in 
the MCMC chain, hence most samples in the chain lie in the region with $R \lesssim 10$. For this reason, the region with $R \lesssim 10$ seems to be 
favored. However, this can be considered as a volume effect in the MCMC analysis.  Therefore the upper bound on $R$ in Fig.~\ref{fig:eps_R_eta0} should be 
interpreted with some caution.

In any case, in the region with large $R$, the spectrum $P_\zeta^{(1)}$ should be subdominant on the CMB scale and
very blue-tilted spectrum can be admissible, which would give large power on small scale. 
Thus, relatively large value of $\mu$ can be obtained in the large $R$ region.

On the other hand, since in the small $R$ region, the tensor-to-scalar ratio is given by the standard formula 
$r = 16\epsilon$, the constraint on $r \lesssim 0.1$ indicates $\epsilon \lesssim 0.006$, which can also be seen 
from Fig.~\ref{fig:eps_R_eta0}.

\bigskip
\bigskip
\noindent
{\bf (iv) Mixed models with $(\eta_1, \eta_2)=(0.2,0)$}

\bigskip
In certain particle physics motivated models of inflation the $\eta$ parameter can turn out to be too large to be consistent with the observed
scale-dependence of the power spectrum.  However, when an additional source of fluctuations exists, the contribution of the inflaton to the power spectrum can be 
subdominant. A large value of $\eta$ can then be reconciled with observations. 

With this kind of a situation in mind, let us here investigate as an example one particular realization by fixing $\eta_1 = 0.2$ and $\eta_2 =0$. 
In the analysis, we vary $R$ and $r$. With the $\eta$ parameters fixed,  $\epsilon$ can be determined through Eq.~\eqref{eq:mixed_r}, and  
the spectral indices $n_{s1}$ and $n_{s2}$ will be completely fixed.

\begin{figure} %[htbp]
  \begin{center}
    \resizebox{160mm}{!}{
    \includegraphics{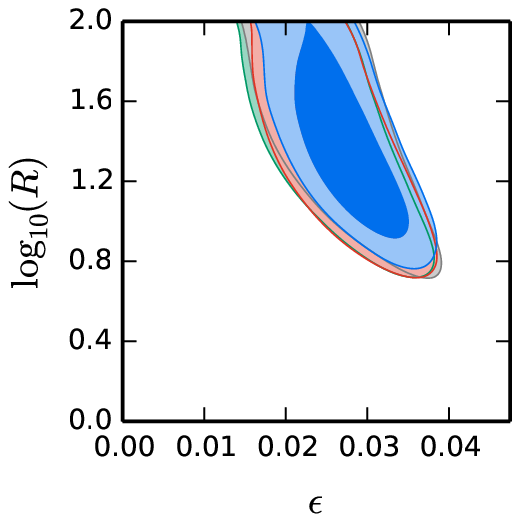}\hspace{3mm}
    \includegraphics{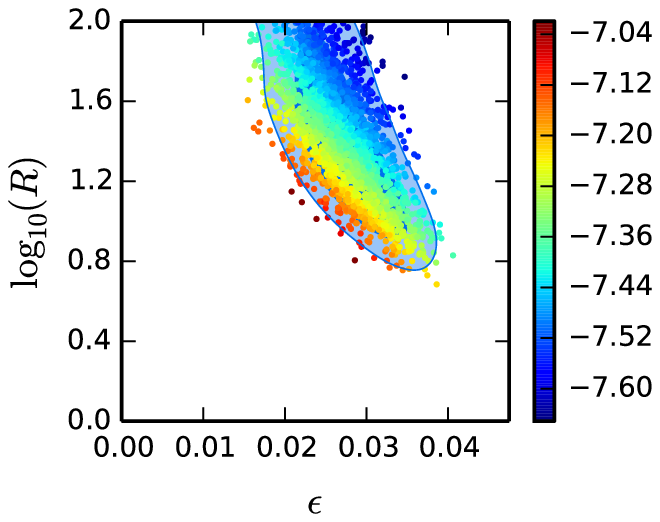}
    }
  \caption{The same as Fig.~\ref{fig:eps_R_mixed}, but for the case with $\eta_1 = 0.2$ and $\eta_2 =0$.}
     \label{fig:eps_R_eta_0_02}
  \end{center}
  \end{figure}

In Fig.~\ref{fig:eps_R_eta_0_02}, we show the CMB constraints and a scatter plot of $\mu$ as in Fig.~\ref{fig:eps_R_mixed}.
Since now $\eta_1$ results in a too blue-tilted power spectrum for $P_\zeta^{(1)}(k)$, it cannot be a dominant source for the
density fluctuations. Thus there exists a lower bound on $R$ and  the data favours the large $R$ region, as is seen from Fig.~\ref{fig:eps_R_eta_0_02}.
When $\eta_1$ is positive and large, the spectral index for $P_\zeta^{(1)}$ becomes highly blue-tilted, which enhances the power on small scales
and gives rise to a large value of $\mu$. This can be seen from the right panel of Fig.~\ref{fig:eps_R_eta_0_02}.
Especially in the allowed region with lower $R$, a relatively large value of $\mu$ tends to be generated because of the large amplitude of $P_\zeta^{(1)}$ with the blue-tilted spectrum.

\bigskip
\bigskip
\noindent
{\bf (v) Mixed models with $\eta_2=0$ and $r= 0.005$}

\bigskip
Currently there are many CMB B-mode polarization experiments that are either on-going or at the planning stage, with an expected
sensitivity to  the primordial gravitational waves at the level of $r = \mathcal{O} (10^{-2})$--$\mathcal{O}(10^{-3})$ 
(see e.g., \cite{Lee:2014cya,Creminelli:2015oda,Huang:2015gca,Errard:2015cxa}).  
To illustrate what would be the implications of the detection of gravitational waves on the mixed models, let us
present an analysis in which the tensor-to-scalar ratio is fixed to be $r=0.005$. We also fix $\eta_2=0$, 
which would be a typical value in mixed inflaton and spectator models. 

\begin{figure} %[htbp]
  \begin{center}
    \resizebox{160mm}{!}{
    \includegraphics{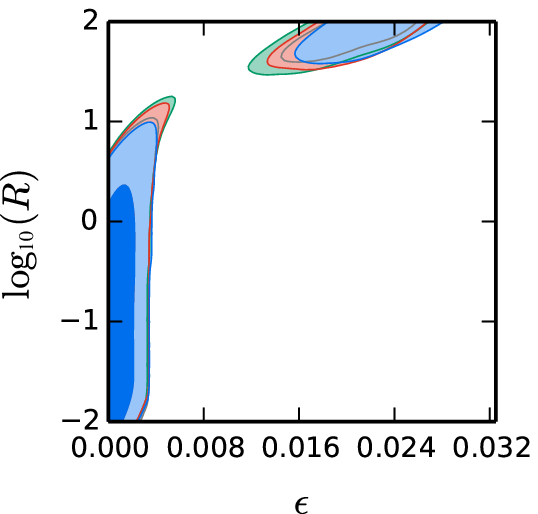}\hspace{3mm}
    \includegraphics{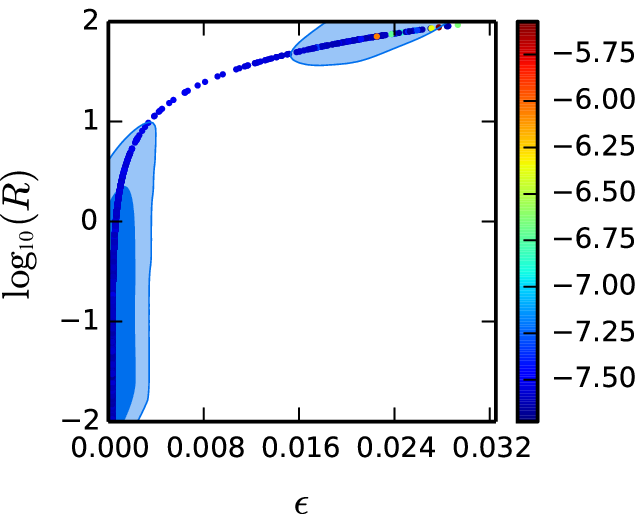}
    }
  \caption{The same as Fig.~\ref{fig:eps_R_mixed}, but for the case with $\eta_2=0$ and $r=0.005$.}
     \label{fig:eps_R_r0.005}
  \end{center}
  \end{figure}

We again show the CMB constraints and a scatter plot of $\mu$ but now for this particular setup in Fig.~\ref{fig:eps_R_r0.005}.
Since the tensor-to-scalar ratio is fixed as $r=0.005$, when $R$ is small (i.e., $P_\zeta^{(1)}$ dominates the power spectrum), 
the $\epsilon$ parameter is fixed as $\epsilon \simeq 3.1 \times 10^{-4}$. Since in this region,  the spectral index $n_{s1}$ is given by 
$n_{s1} -1 = -6 \epsilon + 2 \eta_1$,  the spectral index can be fitted by adjusting $\eta_1$. However, as the value of $R$ increases, 
the contribution from $P_\zeta^{(2)}$ becomes sizable.  Since the spectral index for $P_\zeta^{(2)}$ is given by $n_{s2} - 1 = -2 \epsilon$, 
in the large $R$ regions the spectral index for the total power spectrum cannot be fitted to the data. This is so because the spectral index is mainly 
determined by $\epsilon$, which is also constrained by the fixing of the tensor-to-scalar ratio. 

However, when $R$ becomes very large so that $P_\zeta^{(2)}$ is almost completely responsible for the power spectrum, 
there appears a  ``sweet spot".
Notice that, with the tensor-to-scalar ratio being fixed as $r=0.005$,  the parameters $\epsilon$ and $R$ are related via Eq.~\eqref{eq:mixed_r},
i.e., $\epsilon = 0.005 (1+R)/16$. When $P_\zeta^{(2)}$ is dominant (or $R$ is large), the spectral index is given by 
$n_s -1 = -2 \epsilon = -0.005 (1+R)/8$. Since the spectral index is well determined as $n_s -1 \simeq -0.04$, 
we find that $R\sim 63$ can satisfy the constraint on the spectral index and simultaneously be consistent with $r=0.005$. 
This is the reason why there is an island of the allowed region around $R\sim 63$, as can be seen in Fig.~\ref{fig:eps_R_r0.005}.
As in the case (iii), the value of $\mu$ can be large since, in the large $R$ region,  $n_{s1}$ gives a blue tilt.

\bigskip
\bigskip

\begin{figure} %[htbp]
  \begin{center}
    \resizebox{100mm}{!}{
    \includegraphics{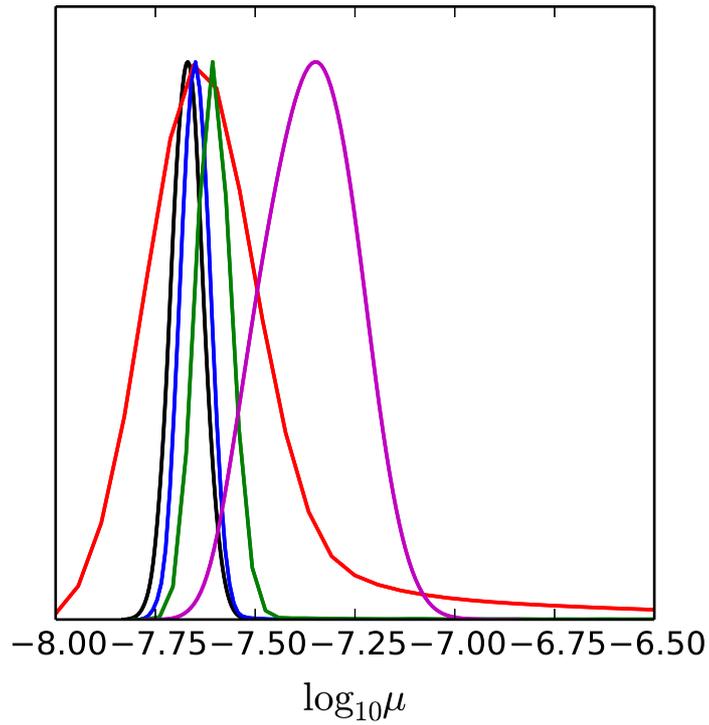}
    }
  \caption{Posterior predictive value for $\mu$ for: the standard single-field inflaton case (black);  the general mixed case (red);
  the mixed case with $(\eta_1, \eta_2) = (0.2,0)$ (purple); the mixed case with $\eta_2 =0$ (blue); and the mixed case with $(r, \eta_2) =(0.005, 0)$ (green). }
     \label{fig:1D_mu }
  \end{center}
  \end{figure}

Finally,  in Fig.~\ref{fig:1D_mu } we show the 1D posterior predictive distribution of $\mu$ for the cases considered above.
Since we have evaluated the value of $\mu$ for each sample (or cosmological parameter set) using the MCMC chains, 
in a sense we may view $\mu$ as a ``derived" parameter.  Thus we can calculate the distribution of $\mu$ for
each MCMC chain, which can be considered as yielding the ``probable" or ``typical" value of $\mu$,  given the current observational constraints. 

From Fig.~\ref{fig:1D_mu }, one can see that the standard single-field inflaton model typically predicts $\mu \sim 2 \times 10^{-8}$ with a  relatively narrow distribution. The expectation is that such values of $\mu$ would be detectable in future experiments.
For the mixed model, the distribution is much broader, and the central value is slightly larger than the one 
in the standard inflation case. This is because the spectral index $n_{s1}$ for the inflaton fluctuations is less constrained within the framework of 
a mixed model; the spectral index can be either very blue-tilted or red-tilted. 
Therefore, a mixed model could easily give rise to a value of $\mu$ that is larger by almost an order of magnitude and therefore would be easily detectable.

When in our example (iv) we assume $\eta_1=0.2$, the spectral index $n_{s1}$ is inevitably blue-tilted. Thus the small scale power is 
enhanced to generate large values of $\mu$, which gives a distribution of $\mu$ that is shifted to larger values as compared to the 
other cases, as can be seen in Fig.~\ref{fig:1D_mu }. Therefore we could conclude that if a large $\mu$ is found in future observations, 
this kind of a particle physics motivated model could become an interesting candidate for the origin of the density fluctuations.

%%%%%%%%%%%%%%%%%%%%%%%
\section{Summary}
\label{sec:summary}
%%%%%%%%%%%%%%%%%%%%%%%

As current cosmological observations impose severe constraints on the spectral index $n_s$ and
the tensor-to-scalar ratio $r$, 
many single-field inflation models have either been excluded or are on the verge of exclusion. In particular, as the sensitivies improve,
a future stringent upper bound on the tensor-to-scalar ratio may
motivate us to take mixed inflaton and spectator field models seriously for the simple reason that $r$ tends to be naturally suppressed in such models (of course, there are also inflaton models with low $r$). 

In mixed models the inflaton and a spectator field can be both responsible for the primordial perturbation so that there are two separate sources, $P_\zeta^{(1)}$and  $P_\zeta^{(2)}$, for the density fluctuations. We have argued that such models could soon be tested, and differentiated from the single-field inflaton
models, by combining the observations on the CMB anisotropies with the data on the $\mu$ distortion of the CMB.

In the present paper, we have investigated in detail the constraints on such models from the current CMB observations 
 by adopting an MCMC analysis.  We note that since it is possible that
$P_\zeta^{(1)}$ gives the dominant contribution to the large scale fluctuations while $P_\zeta^{(2)}$ contributes to the small scales, or vice versa, 
a very blue-tilted spectral index for the power spectrum subdominant on CMB scales could be allowed for. We also point out that
such possibilities could easily be ruled out in future determinations of the $\mu$ distortion in experiments such as PIXIE and PRISM, which are expected to be able to detect $\mu$ distortion at the level of $\mu \sim  10^{-8}- 10^{-9}$. In 
Fig.~\ref{fig:1D_mu } we demonstrate that the standard single-field inflaton model typically predicts $\mu \sim 2 \times 10^{-8}$ with a  relatively narrow distribution, whereas in mixed models the distribution can be very broad.  Indeed, 
a mixed model could easily give rise to $\mu \sim  10^{-7}$ and be thereby easily distinguishable from the single-field inflaton case.

In addition to the general mixed model, we also analyzed certain specific representative cases by fixing the values of $\eta_1$ and/or $\eta_2$. 
For example, in some inflationary models $\eta$ can be large with $|\eta| \sim \cal{O}$(0.1), which in the standard single-field inflation is not admissible because the spectral index becomes either too red-tilted or too blue-tilted.
However, in mixed models, such a possibility can be realized provided the scalar field with large $\eta$ does not affect 
large scale density fluctuations, which could be sourced mainly by the other scalar. In the future, such models can be confronted with the
small scale observations of the $\mu$ distortion. 

Our study, while remaining incomplete in that we do not perform a systematic survey of all possible models, nevertheless suggests that one should check  the spectral predictions of a given model over a large range of scales. In this regard the $\mu$ distortion appears to provide a promising new tool for testing critically
models of the primordial perturbation.

\section*{Acknowledgments}

TT would like to thank the Helsinki Institute of Physics for the hospitality
during the visit, where this work was initiated.
The work of TT  is partially supported by JSPS KAKENHI Grant Number 15K05084  and 
MEXT KAKENHI Grant Number 15H05888. 
This work was supported by IBS under the project code, IBS-R018-D1.
We thank the CSC - IT Center for Science (Finland) for computational resources.

\end{document}